\documentclass[pop,reprint,showpacs,groupedaddress,amsmath,amssymb]{revtex4-1}

\usepackage[dvips]{graphicx}

\newcommand{\beq}{\begin{equation}}
\newcommand{\eeq}{\end{equation}}
\newcommand{\bea}{\begin{eqnarray}}
\newcommand{\eea}{\end{eqnarray}}
\newcommand{\mbf}[1]{\mathbf{#1}}
\newcommand{\mrm}[1]{\mathrm{#1}}
\newcommand{\epsi}{\varepsilon}
\newcommand{\del}{\nabla}
\newcommand{\divr}{\nabla \cdot}

\newcommand{\dsub}[1]{\partial_{#1}}
\newcommand{\dsup}[1]{\partial^{#1}}
\newcommand{\wt}[1]{\widetilde{#1}}
\DeclareMathOperator{\dalem}{\square^2}
\DeclareMathOperator{\expi}{\exp\!i}
\newcommand{\abs}[1]{{\lvert {#1} \rvert}}

\newcommand{\uvec}[1]{{\hat{\mbf{#1}}}}

\begin{document}


\title{Vector potential analysis of the helicon antenna in vacuum} 



\author{Robert W. Johnson}
\email[]{robjohnson@alphawaveresearch.com}
\homepage[]{http://www.alphawaveresearch.com}
\affiliation{Alphawave Research, Jonesboro, GA 30238, USA}


\date{\today}

\begin{abstract}
The helicon antenna is a well-known device in the field of electric propulsion.  Here we investigate the vector potential produced in vacuum by such an antenna with typical size parameters.  Both a static and a dynamic analysis are performed.  The dynamic calculation is evaluated at both the usual operating frequency and one which is slightly greater.  At the higher frequency, a pulse of electromagnetic energy is found to propagate along the cylindrical axis in either direction.  The possible adaptation of the helicon antenna as a RF injection device for burning plasma is discussed.
\end{abstract}


\maketitle 


\section{Introduction}
The helicon antenna, consisting of two rings separated and joined by two curving limbs, is a well-known device in its application as an ion source for electric propulsion~\cite{chen-137,yano-063501,palmer-0609,chen-123507,arefiev-062107}.  Driven by a radio-frequency amplifier, its energy couples to the plasma medium to drive the dissociation of propellant material, such as argon gas.  Here we investigate the vector potential produced by such a device in vacuum, with an eye towards its adaptation as a source of RF energy for injection into a burning plasma.

Our methodology is based on the numerical evaluation of solutions to the inhomogeneous field equations $\dalem A^\mu = - \mu_0 J^\mu$ in Lorenz gauge $\dsub{\mu} A^\mu = 0$, where $A^\mu \equiv (a/c_0, \mbf{A})$ is the four-vector for the potential, $J^\mu \equiv (c_0 j, \mbf{J})$ is the four-vector for the current source, and $c_0^{-2} \equiv \mu_0 \epsi_0$ gives the speed of light in vacuum.  The d'Alembertian operator is $\dalem \equiv \dsup{\nu} \dsub{\nu} = - c_0^{-2} \dsub{t}^2 + \del^2$ in a frame with metric signature $(-, +, +, +)$.  We subscribe to the reductionist viewpoint~\cite{rousseaux-2005-30,rouss-00013234} that the potential is a more fundamental description than the field formulation, and that the Lorenz gauge makes explicit the relation between the continuity of the potential and the continuity of the source $\dsub{\mu} J^\mu = 0$.

The solution $A^\mu$ for a given $J^\mu$ is found by directly imaging the current source throughout the region, by which we mean that the contribution from each source element $J^\mu(t',\mbf{r}')$ is calculated at a point $(t,\mbf{r})$ with regard to the propagation delay, or retarded time $t' = t - \Delta_r / c_0$, as well as the spatial distance to the element $\Delta_r = \abs{\mbf{r} - \mbf{r}'}$.  The antenna is assumed to be driven by a feed current arranged so that its contribution is negligible, leaving to calculate only the potential produced by the antenna.  We neglect any charge accumulation along the conductor, so that the static charge density vanishes, thus $j=0$ and $\divr \mbf{J} = 0$.  The solution to the wave equation for a line source is then written
\beq \mbf{A}_C (t,\mbf{r}) = \dfrac{\mu_0}{4 \pi} \int_{\mbf{r}'} \dfrac{\mbf{I}_C (t',\mbf{r}')}{\Delta_r} dl  \;, \eeq
where the subscript $C$ reminds us that these vectors must be expressed in Cartesian components $\mbf{A} = (A_X, A_Y,A_Z)$, neglecting any self-interaction between different parts of the antenna.

The antenna is modeled as a collection of discrete line current elements $\mbf{I}_k \equiv \mbf{I} (\mbf{r}_k')$, so that the integral becomes a sum $\int_{\mbf{r}'} dl \rightarrow \sum_k \Delta_k$, where $\Delta_k$ is the length of the $k$th element.  The antenna is assigned typical size parameters of half-width $w=0.1$ m and half-height $h=0.2$ m (or 1 and 2 dm respectively), and the feed current is normalized to $\mu_0 I_0 / 4 \pi \equiv 1$.  We assume that the driving current is provided along a twisted pair cable so that the opposing phases produce canceling contributions to the overall potential.  The Cartesian $(X, Y, Z)$ and cylindrical $(R, \phi, Z)$ coordinates are related by $X = R \cos \phi$ and $Y = R \sin \phi$.  The static analysis calculates the potential in the $XY$ plane at various $Z$ along the cylindrical axis.  The dynamic analysis calculates the potential in the $ZR$ plane at various angles $\phi$ around the cylindrical axis.  The propagation analysis computes $\mbf{A}$ in the $ZR$ plane at $\phi=\pi/4$ and various times $t$.  Our methodology is a straightforward implementation of classical electrodynamics in the potential formulation, such as found in Ref.~\cite{griffiths-89}.

\begin{figure}[]
\includegraphics[width=8.5cm]{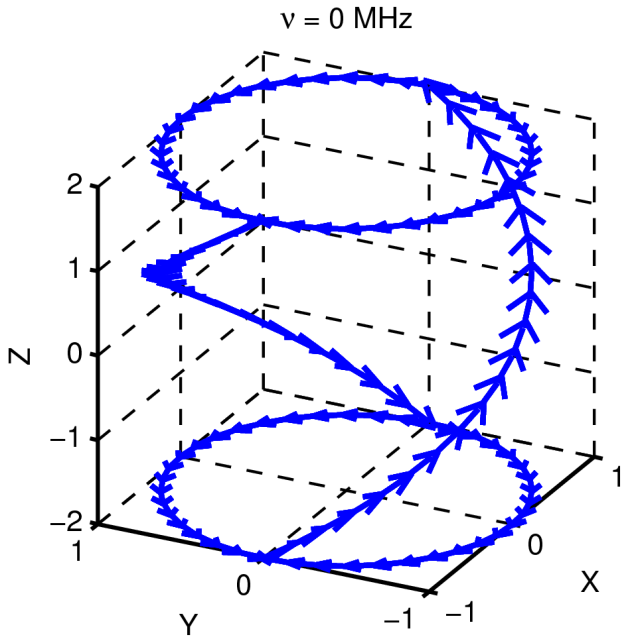}%
\caption{\label{fig:qsaA} The helicon antenna is modeled as a collection of discrete line current elements as described in the text.  The geometry is shown in units of 1 dm = 0.1 m throughout.}
\end{figure}

\section{Static Analysis}
For the static, time independent analysis, the Lorenz gauge is equivalent to the Coulomb gauge, $\divr \mbf{A}=0$.  Position $\mbf{r}'(\lambda,\phi')$ along the helicon antenna may be parametrized by the azimuth $\phi'$ and the line distance along the conductor $\lambda$ to the location where the driving current is attached, here taken to be at $(X', Y', Z') = (w, 0, -h)$.  The ring at $Z' = h$ will be called the ``head'', and the one at $Z' = -h$ will be the ``tail''.  The vertical limbs we call ``arms'', and the geometry of the antenna is depicted in Fig.~\ref{fig:qsaA}, where the ``legs'' carrying the driving current are not shown.  It takes two revolutions around the circle to cover the entire antenna, thus the positive half is identified with $\phi_+' \in [0, 2 \pi]$ and the negative with $\phi_-' \in [0, -2 \pi]$.  The azimuth is parametrized over $N_\phi$ values, so that $\Delta_\phi = 2 \pi / N_\phi$.  The line elements for the rings have a length $\Delta_k = w \Delta_\phi$, and those for the arms have a length $\Delta_k = ( \Delta_Z^2 + w^2 \Delta_\phi^2 )^{1/2}$, where $\Delta_Z = 2 h \Delta_\phi / \pi$ is the vertical extent for arm elements at $Z_k' = 2 h \phi_k' / \pi - h$.  Alternately, one could take $\sum_k \rightarrow \sum_{\phi'} \sum_{\mathrm{head, tail, arm}}$ for one revolution of the circle $\phi' \in [0, 2 \pi]$.

By Kirchhoff's point rule, the magnitude of the current along either branch of each ring is $I_r = I_0/2$, and along the arms it is $I_a = I_0$.  The arms must cover a height of $2 h$ over an azimuthal range of $\pi$, thus the ratios  $I_Z / I_\phi = 2 h / \pi w$ and $(I_a/I_\phi)^2 = 1 + (I_Z / I_\phi)^2$ determine $I_X = - I_\phi \sin \phi_k'$ and $I_Y = I_\phi \cos \phi_k'$ for $I_\phi = I_0 [1 + (2 h / \pi w)^2]^{-1/2}$ so that $\mbf{I}_a = I_0 (- \sin \phi_k', \cos \phi_k', 2 h / \pi w)$, and similarly for $\mbf{I}_r$ with sign changes as appropriate for the two branches of each ring.  The contribution from each element $\mbf{I} (\mbf{r}_k')$ to the solution $\mbf{A} (\mbf{r})$ is calculated by multiplying the current $\mbf{I}_k$ by $\Delta_k / \Delta_r$ for each component in $(X, Y, Z)$.

\begin{figure}[]
\includegraphics[width=8.5cm]{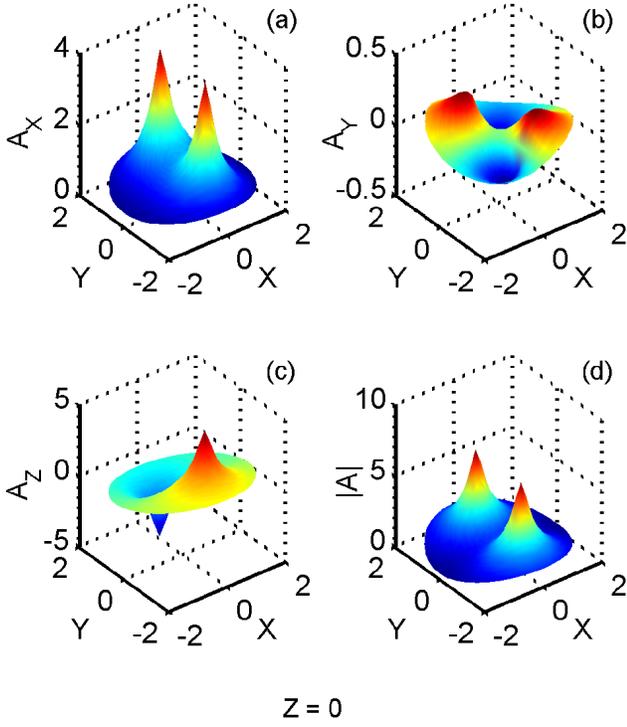}%
\caption{\label{fig:qsaB} Static vector potential at $Z=0$.}
\end{figure}

\begin{figure}[]
\includegraphics[width=8.5cm]{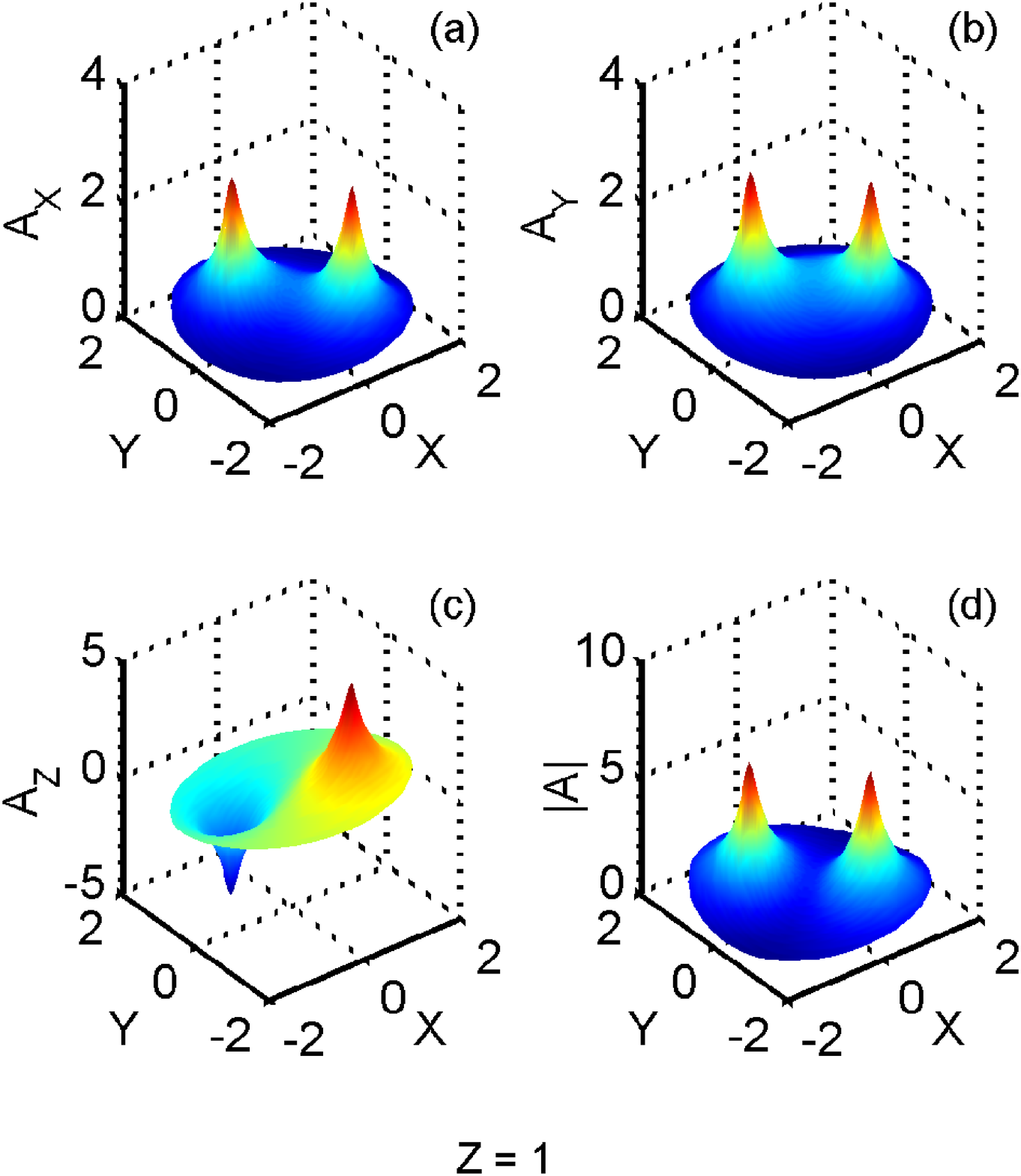}%
\caption{\label{fig:qsaC} Static vector potential at $Z=1$.}
\end{figure}

\begin{figure}[]
\includegraphics[width=8.5cm]{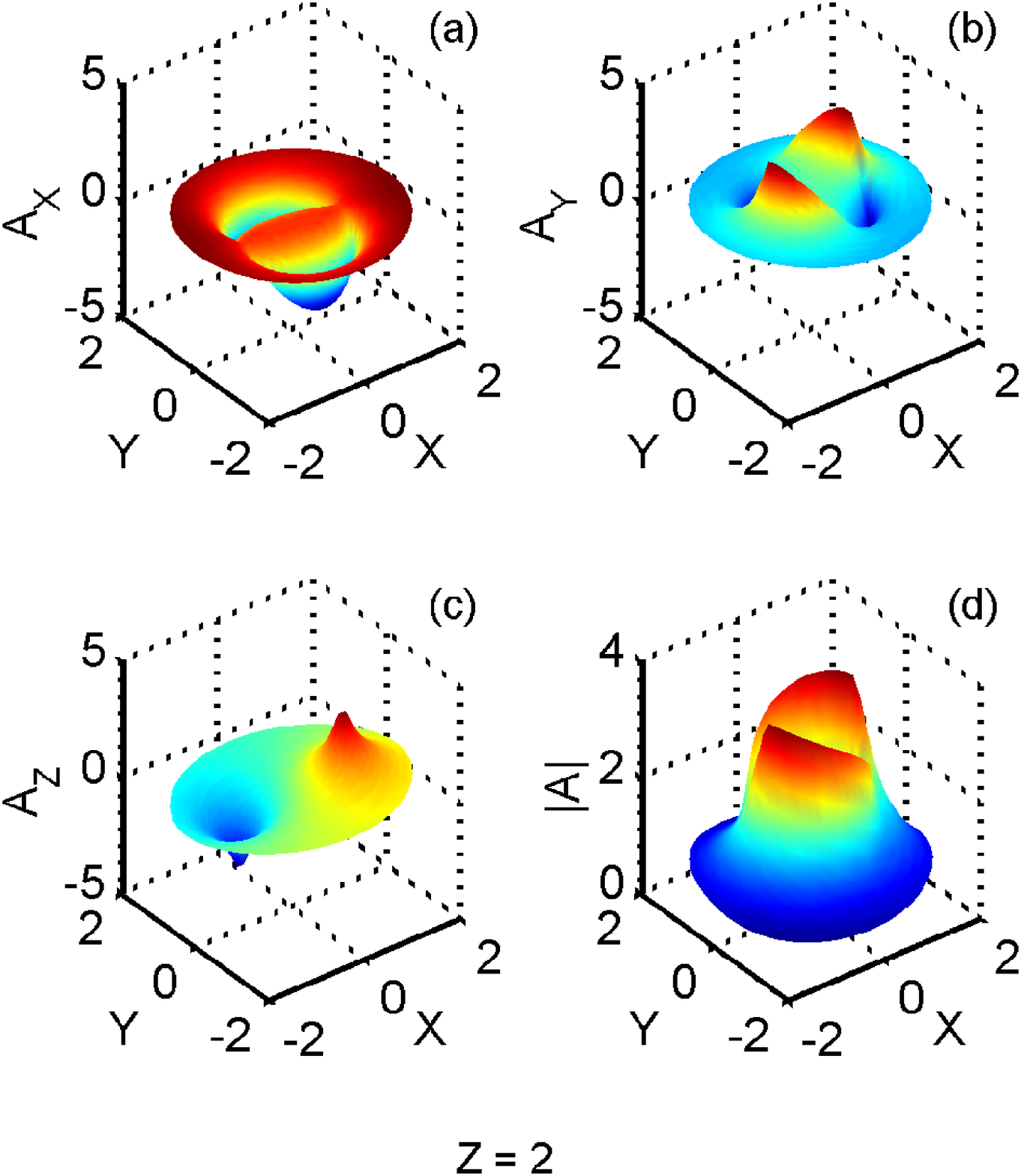}%
\caption{\label{fig:qsaD} Static vector potential at $Z=2$.}
\end{figure}

\begin{figure}[]
\includegraphics[width=8.5cm]{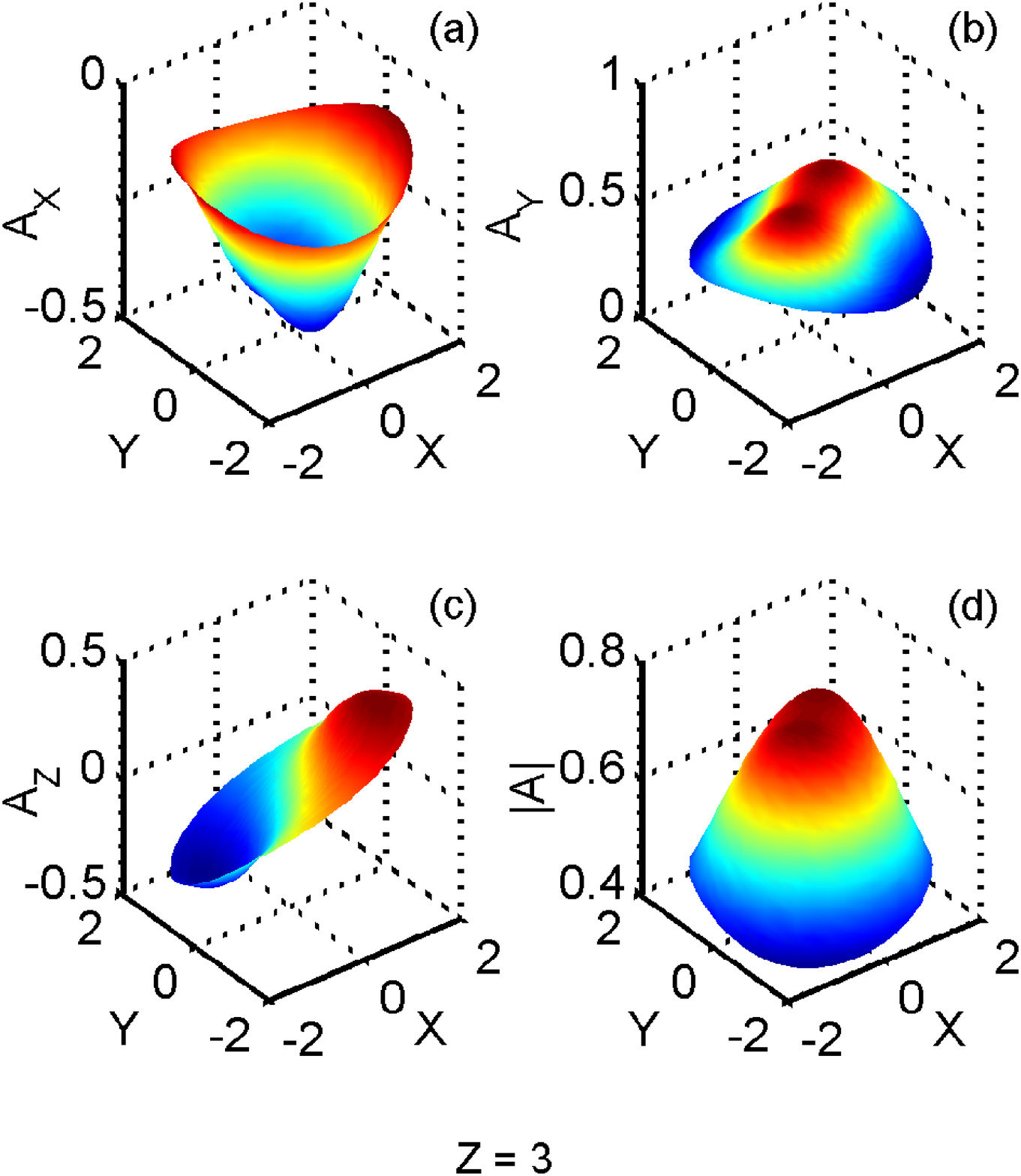}%
\caption{\label{fig:qsaE} Static vector potential at $Z=3$.}
\end{figure}

On the midplane of the antenna $Z=0$ dm, the vector potential $\mbf{A}$ has its greatest magnitude at the locations of the arm elements $(0, \pm w, 0)$, as seen in Fig.~\ref{fig:qsaB}, and is dominated by the $X$ and $Z$ components as expected; the $Y$ component is about a tenth of the magnitude of the others.  At three quarters of the antenna height, $Z=1$ dm shown in Fig.~\ref{fig:qsaC}, the locations of the arms are clearly discerned in the vector potential.  In the plane of the head ring, $Z=2$ dm shown in Fig.~\ref{fig:qsaD}, the vector potential is mostly in the azimuthal direction, with a strong vertical component only at the junctions of the head ring with the arms.  Away from the antenna, at $Z=3$ dm shown in Fig.~\ref{fig:qsaE}, a significant vector potential still exists, with both vertical and azimuthal components.

\section{Dynamic analysis}
Let us now consider driving the antenna with an oscillatory current of frequency $\nu = 1 / \tau > 0$, whose wavelength in vacuum is $\lambda_0 = c_0 / \nu$.  The phase of the current along the antenna is determined by the line distance along the conductor $\lambda (\phi')$ from the current element to the location selected as the zero phase reference, where the feed current attaches to the antenna.  The antenna has a half-length of $\lambda_\mrm{max} = \lambda (\pm 2 \pi) = \pi w + [(\pi w)^2 + (2 h)^2]^{1/2}$.  The way we have orientated our antenna, the positive branch consists of the tail ring and the arm with $Y<0$, while the negative branch is the head ring and the arm with $Y>0$.  To reduce the number of brackets, let us denote $\expi (\alpha) \equiv \exp (i \alpha)$.  The complex current along the positive branch of the conductor is $\wt{I}_+ (t, \lambda) = I_0 \expi [2 \pi (\lambda / \lambda_0 - t / \tau)] = \wt{I}_0 (t) \expi (\theta_\lambda)$, where $\wt{I}_0 (t) = I_0 \expi (\theta_t)$ for $\theta_t = - 2 \pi t / \tau$ and $\theta_\lambda = 2 \pi \lambda / \lambda_0$.  Along the negative branch of the conductor, the current is $\wt{I}_- (t, \lambda) = -\wt{I}_+ (t, \lambda)$.  Propagation delay introduces the retarded phase $\theta_r = 2 \pi \Delta_r / \lambda_0$ such that $\wt{I}_+ (t', \lambda) = \wt{I}_0 \expi (\theta_\lambda + \theta_r)$.  Here, $I_0$ equals $I_r$ or $I_a$ as appropriate.  The complex vector potential $\wt{\mbf{A}}_k (t, \mbf{r})$ is calculated by imaging each discrete element $\wt{\mbf{I}} (t, \mbf{r}_k')$ throughout the region with weighting $\Delta_k / \Delta_r$ and net phase $\expi (\theta_k + \theta_t + \theta_r)$ for $\theta_k = 2 \pi \lambda_k / \lambda_0$, so that the physical potential is $\mbf{A} = \mrm{Re}\; \sum_k \wt{\mbf{A}}_k$.

\subsection{Low frequency}
\begin{figure}[]
\includegraphics[width=8.5cm]{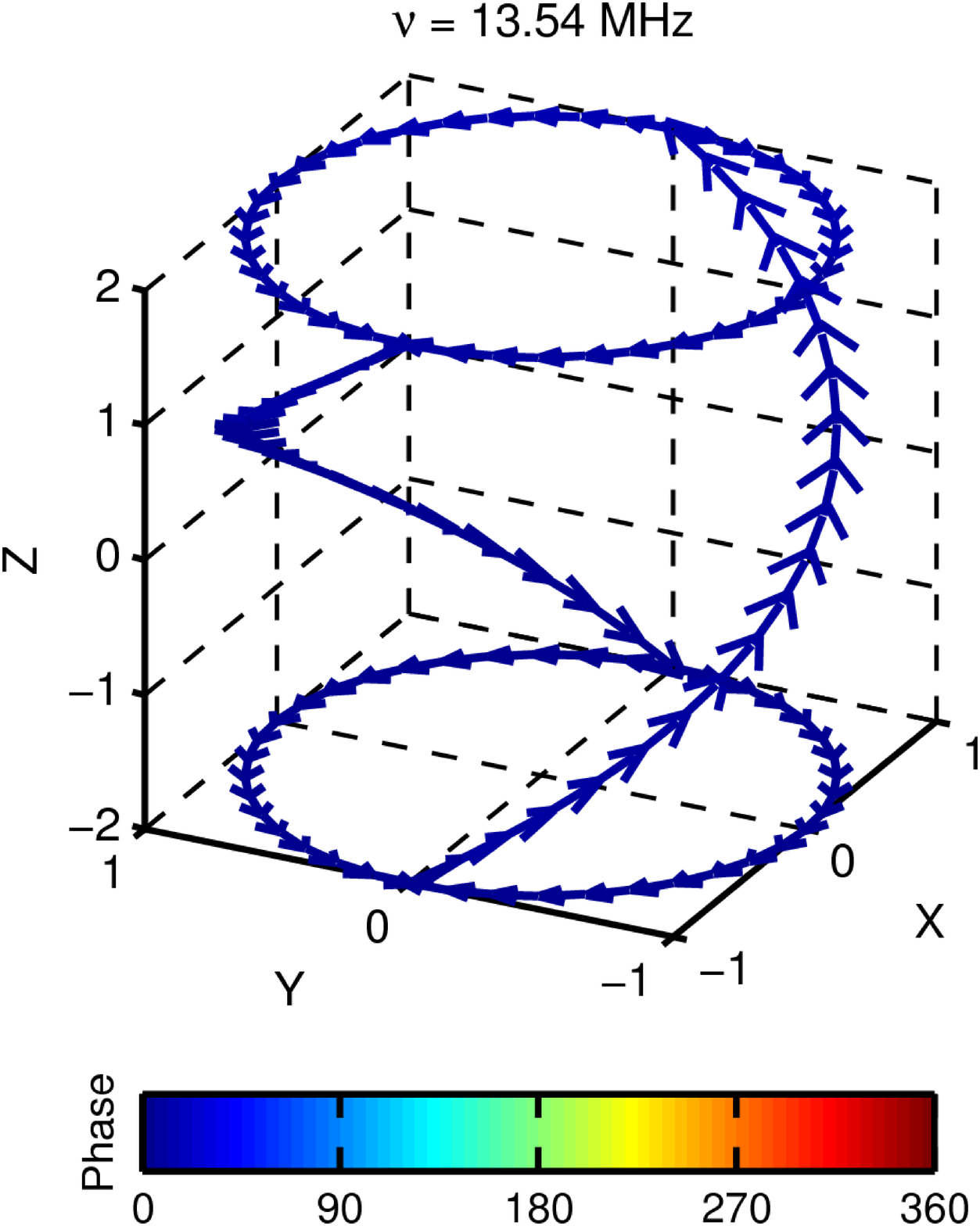}%
\caption{\label{fig:retA} For a driving frequency of $\nu = 13.54$ MHz, the phase difference at the maximum distance $\lambda_\mrm{max}$ from the location of the current feed legs is 3.7\%.  The colorbar indicates the phase in units of degrees, and the length of the arrow represents the magnitude of the real current at $t / \tau = 0$.}
\end{figure}

Considering a driving frequency of $\nu = 13.54$ MHz commonly used to excite argon gas, the vacuum wavelength is $\lambda_0 \sim 22$ m, and $\lambda_\mrm{max} / \lambda_0 = 3.7$\%.  The first step is to calculate the phase of the current along the antenna $\theta_k$, which we show in Fig.~\ref{fig:retA}.  The current is then expressed in Cartesian coordinates $\mbf{I} = (I_X, I_Y, I_Z)$ for each element at $\mbf{r}_k'$ with phase $\theta_k$.  Next one needs the distances $\Delta_r (\mbf{r})$ throughout the region from $\mbf{r}_k'$ to $\mbf{r}$.  From $\Delta_r$ one gets the magnitude weighting $\Delta_k / \Delta_r$ and the retarded phase $\theta_r$.  We set $t / \tau=0$ so that we can investigate the spatial distribution of the vector potential in the $ZR$ plane at various azimuths $\phi$.

\begin{figure}[]
\includegraphics[width=8.5cm]{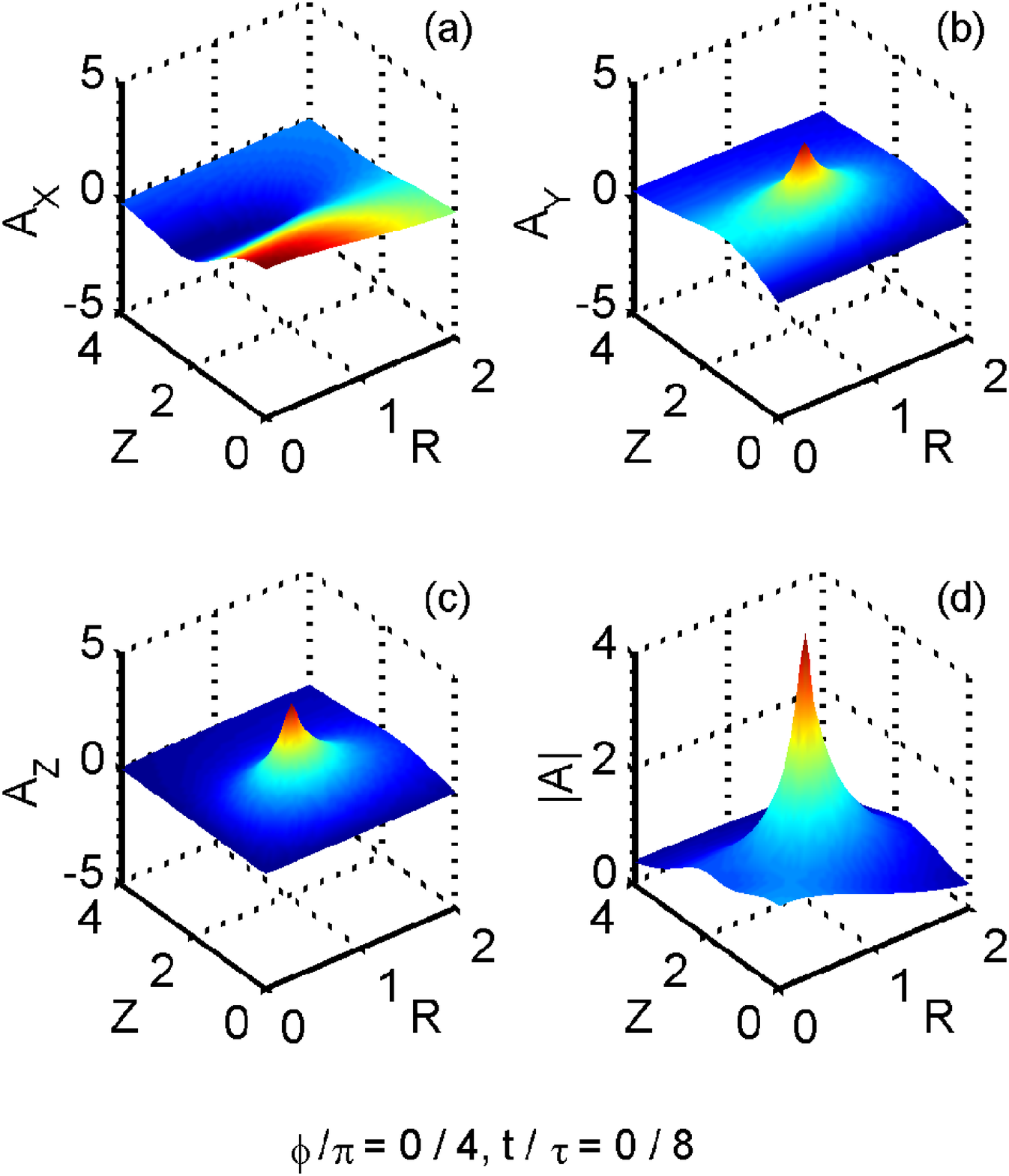}%
\caption{\label{fig:retB} Vector potential for $\nu = 13.54$ MHz at $\phi = 0$ and $t / \tau = 0$.}
\end{figure}

\begin{figure}[]
\includegraphics[width=8.5cm]{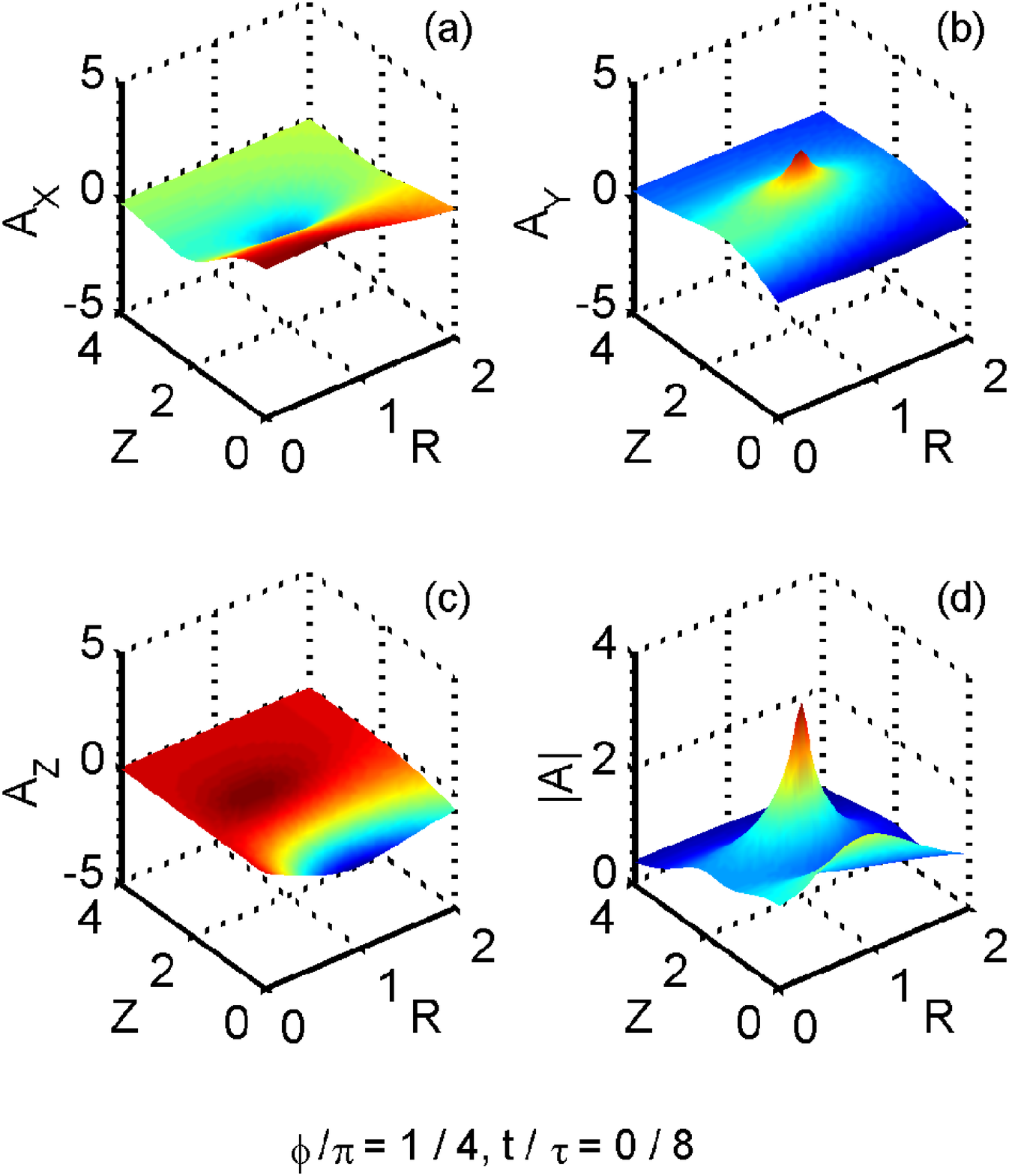}%
\caption{\label{fig:retC} Vector potential for $\nu = 13.54$ MHz at $\phi = \pi / 4$ and $t / \tau = 0$.}
\end{figure}

\begin{figure}[]
\includegraphics[width=8.5cm]{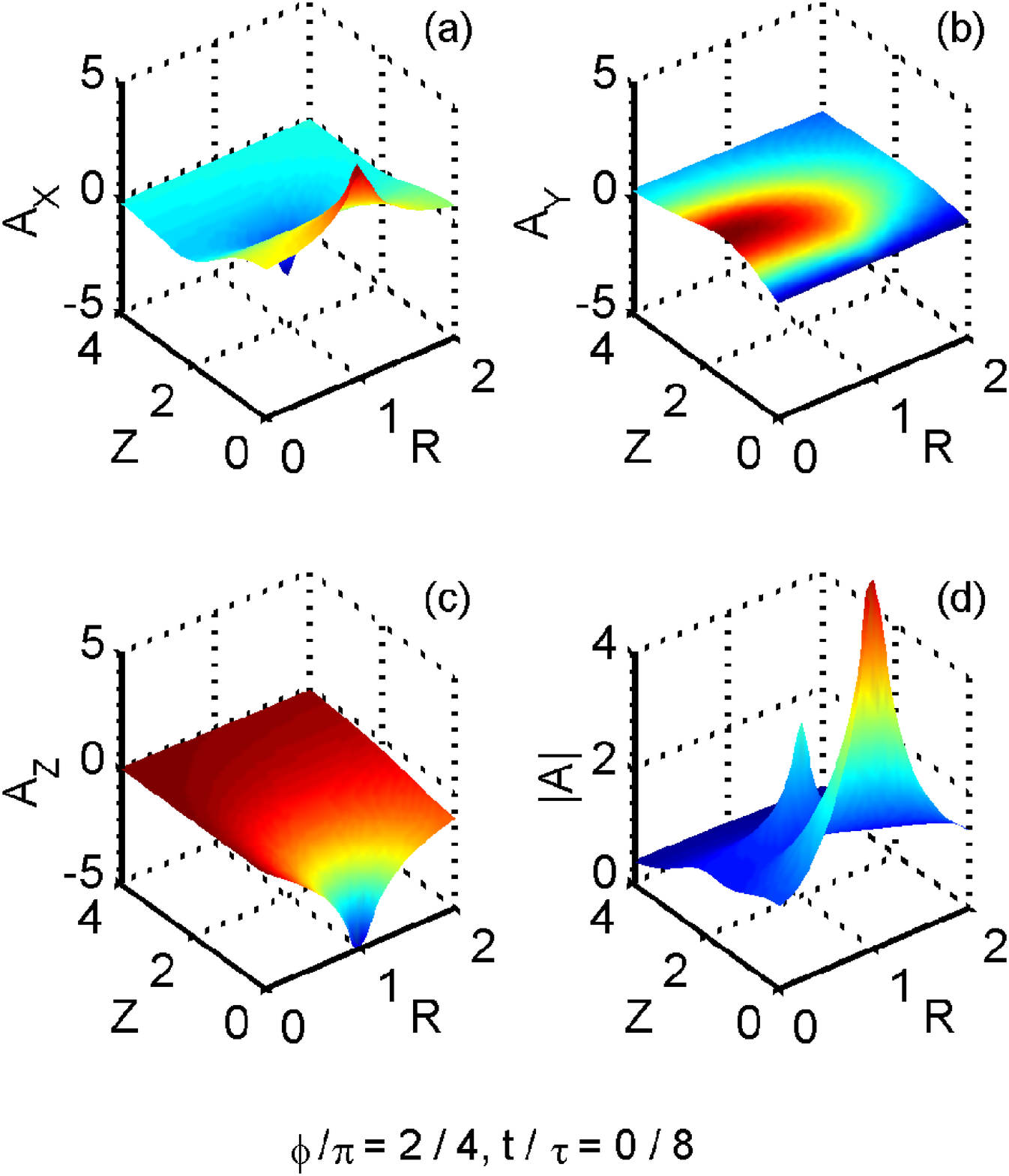}%
\caption{\label{fig:retD} Vector potential for $\nu = 13.54$ MHz at $\phi = \pi / 2$ and $t / \tau = 0$.}
\end{figure}

\begin{figure}[]
\includegraphics[width=8.5cm]{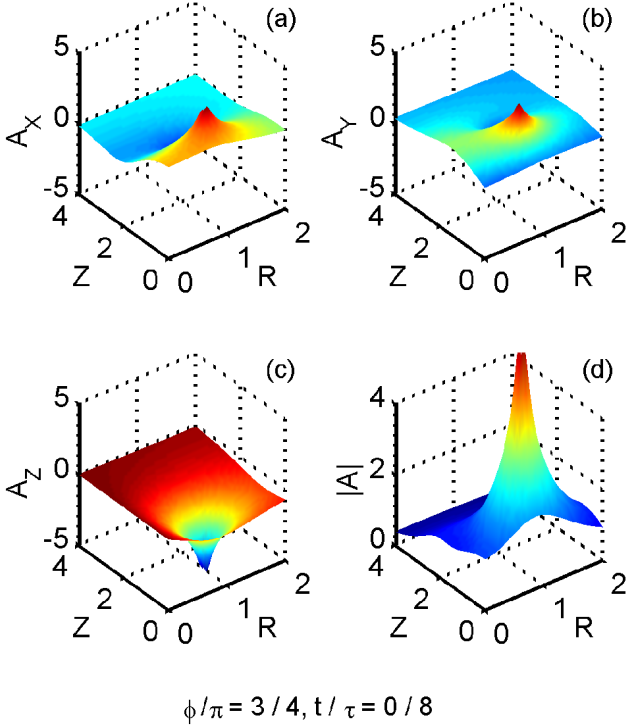}%
\caption{\label{fig:retE} Vector potential for $\nu = 13.54$ MHz at $\phi = 3 \pi / 4$ and $t / \tau = 0$.}
\end{figure}

Along the $X$ axis, where $\phi=0$, the vector potential has the structure for $Z, R > 0$ shown in Fig.~\ref{fig:retB}.  The magnitude $\abs{\mbf{A}}$ is dominated by contributions in the $Y$ and $Z$ directions at the location of the junction of the arm and the head ring, with a component in the $X$ direction for the interior of the antenna $R < w$.  In the plane at $\phi = \pi / 4$ in Fig.~\ref{fig:retC}, the strongest contribution is along $\uvec{Y}$ at the height of the head ring.  On the $Y$ axis, $\phi = \pi / 2$ in Fig.~\ref{fig:retD}, the potential is mostly in the $X$ and $Z$ directions, with a component in the $Y$ direction for the antenna interior.  At $\phi = 3 \pi / 4$ in Fig.~\ref{fig:retE}, the potential has a strong $Z$ component from the antenna arm, with some additional structure along $\uvec{X}$ and $\uvec{Y}$.

\clearpage

\subsection{High frequency}
\begin{figure}[]
\includegraphics[width=8.5cm]{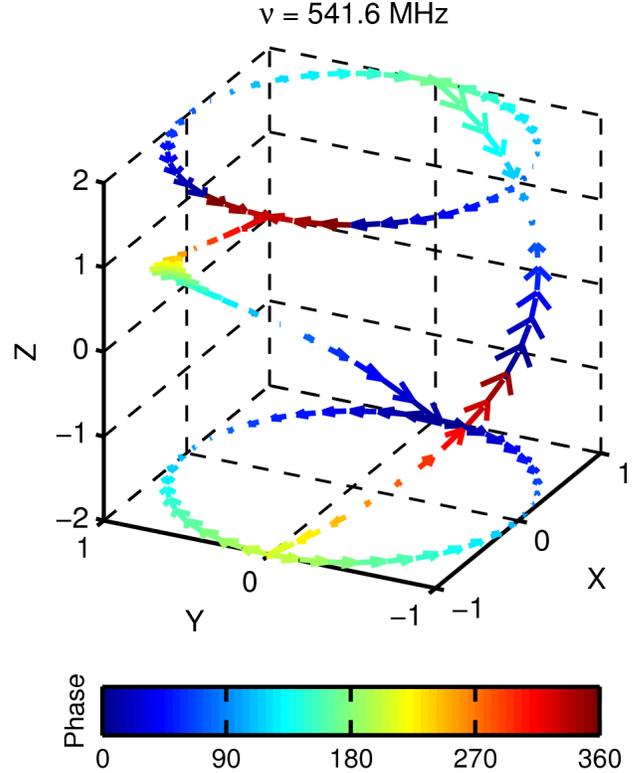}%
\caption{\label{fig:retF} For a driving frequency of $\nu = 541.6$ MHz, the phase difference at the maximum distance $\lambda_\mrm{max}$ from the location of the current feed legs is 149\%.  The colorbar indicates the phase in units of degrees, and the length of the arrow represents the magnitude of the real current at $t / \tau = 0$.}
\end{figure}

\begin{figure}[]
\includegraphics[width=8.5cm]{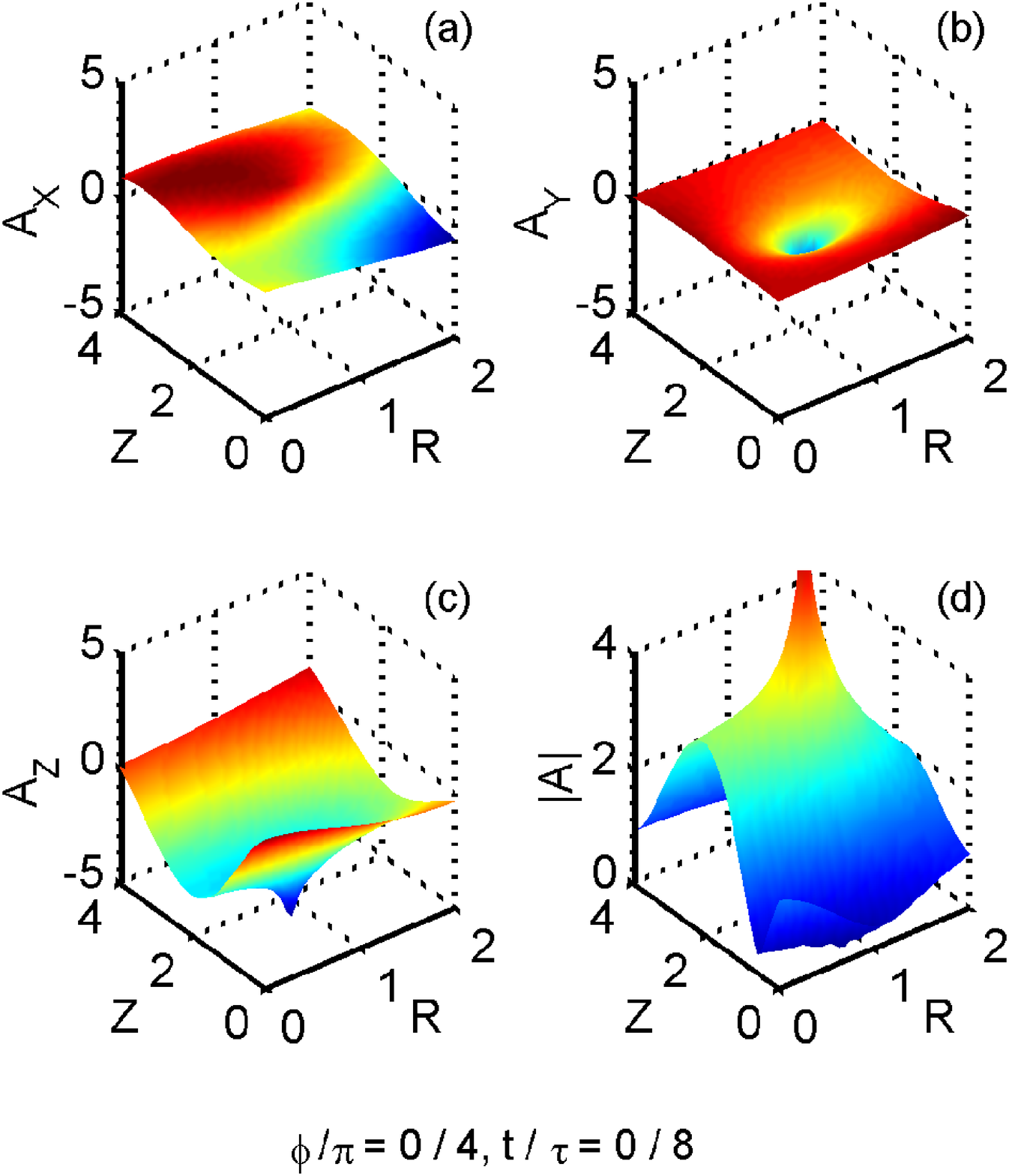}%
\caption{\label{fig:retG} Vector potential for $\nu = 541.6$ MHz at $\phi = 0$ and $t / \tau = 0$.}
\end{figure}

\begin{figure}[]
\includegraphics[width=8.5cm]{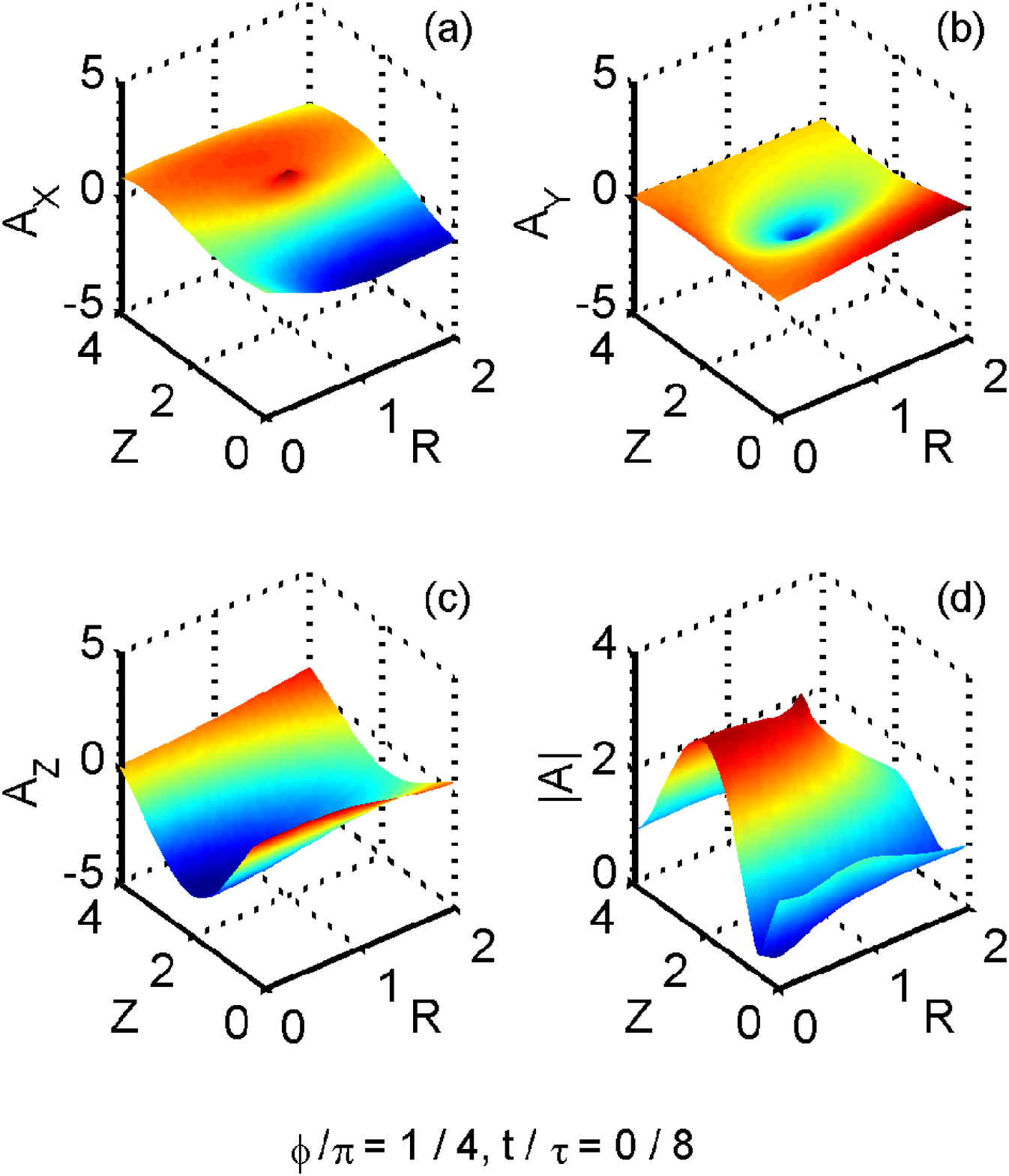}%
\caption{\label{fig:retH} Vector potential for $\nu = 541.6$ MHz at $\phi = \pi / 4$ and $t / \tau = 0$.}
\end{figure}

\begin{figure}[]
\includegraphics[width=8.5cm]{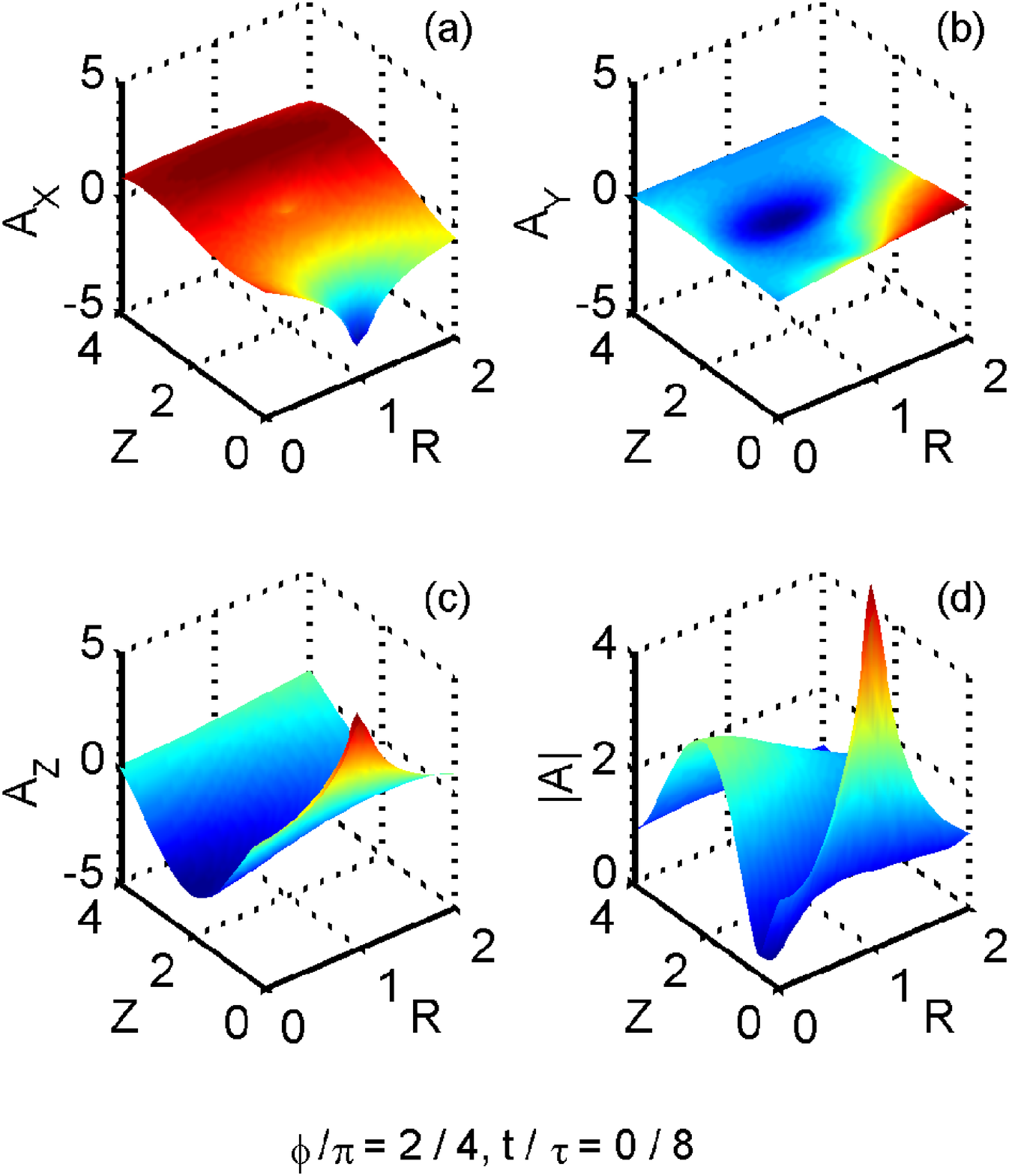}%
\caption{\label{fig:retI} Vector potential for $\nu = 541.6$ MHz at $\phi = \pi / 2$ and $t / \tau = 0$.}
\end{figure}

\begin{figure}[]
\includegraphics[width=8.5cm]{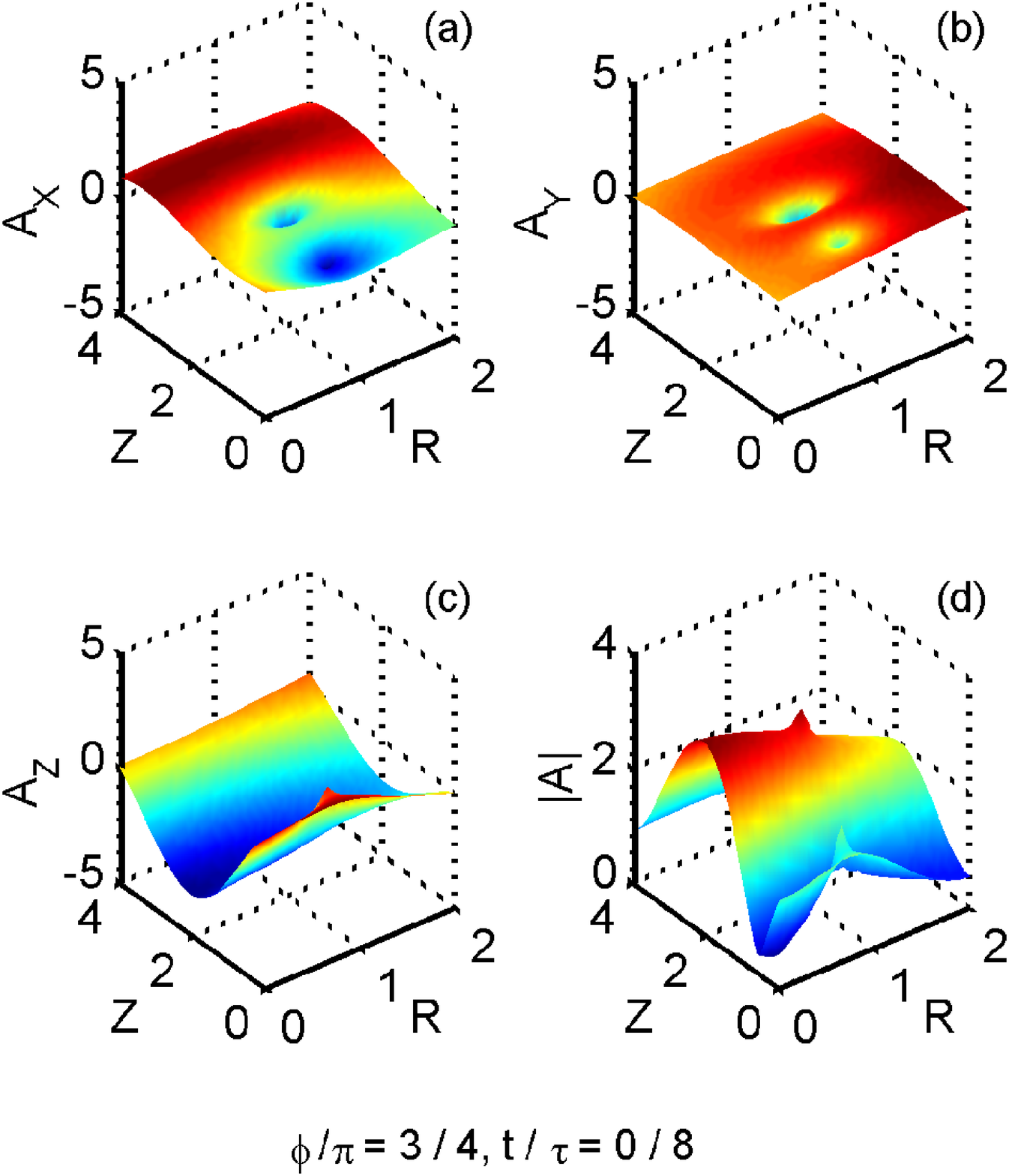}%
\caption{\label{fig:retJ} Vector potential for $\nu = 541.6$ MHz at $\phi = 3 \pi / 4$ and $t / \tau = 0$.}
\end{figure}

Now let us suppose that the antenna is driven at a higher frequency, say $\nu = 541.6$ MHz, whose vacuum  wavelength is considerably shorter $\lambda_0 \sim 0.55$ m so that $\lambda_{max} / \lambda_0 = 149$\%.  The phase structure of the current along the antenna is displayed in Fig.~\ref{fig:retF}, where we see that the magnitude of the real current is zero where $\theta_k + \theta_t = \pm \pi / 2$.  Repeating the azimuthal scan at the same angles as above for $t / \tau = 0$, the results are displayed in Fig.~\ref{fig:retG} through Fig.~\ref{fig:retJ}.  The vector potential has a complicated structure, the most interesting feature of which is the strong magnitude of $\mbf{A}$ which develops at azimuths $\phi = \pi / 4$ and $\phi = 3 \pi / 4$.

\begin{figure}[]
\includegraphics[width=8.5cm]{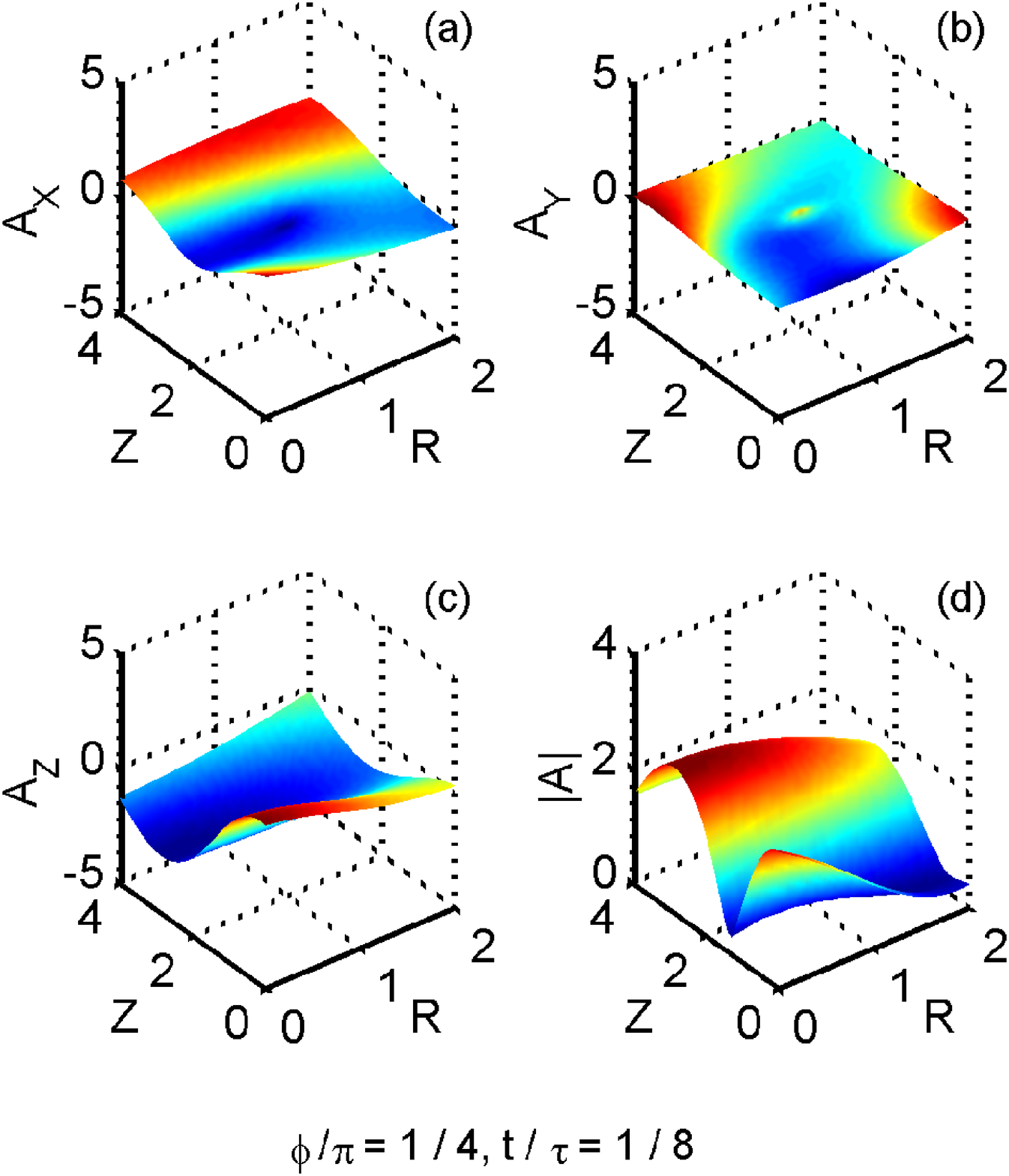}%
\caption{\label{fig:retM} Vector potential for $\nu = 541.6$ MHz at $\phi = \pi / 4$ and $t / \tau = 1/8$.}
\end{figure}

\begin{figure}[]
\includegraphics[width=8.5cm]{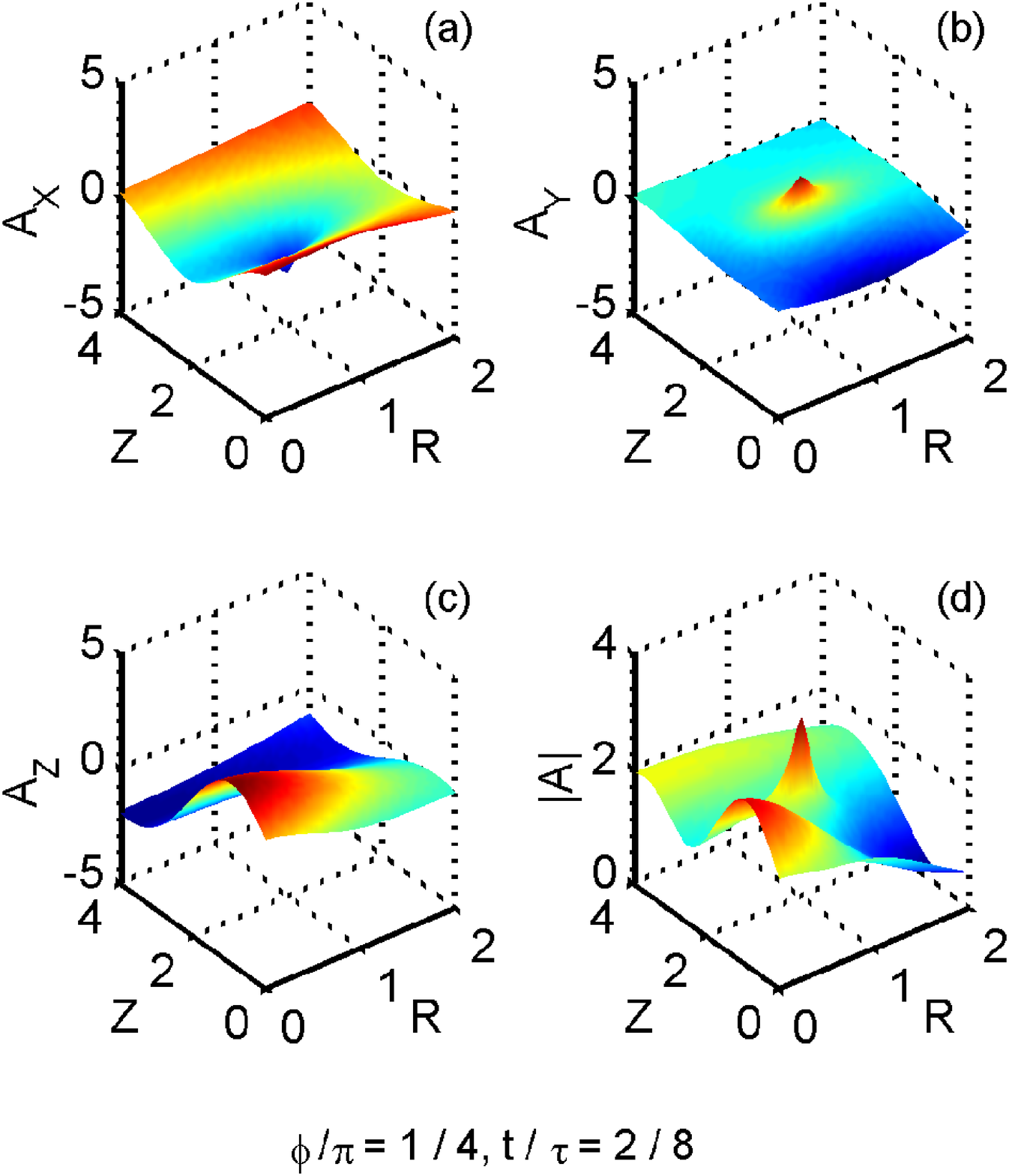}%
\caption{\label{fig:retN} Vector potential for $\nu = 541.6$ MHz at $\phi = \pi / 4$ and $t / \tau = 1/4$.}
\end{figure}

\begin{figure}[]
\includegraphics[width=8.5cm]{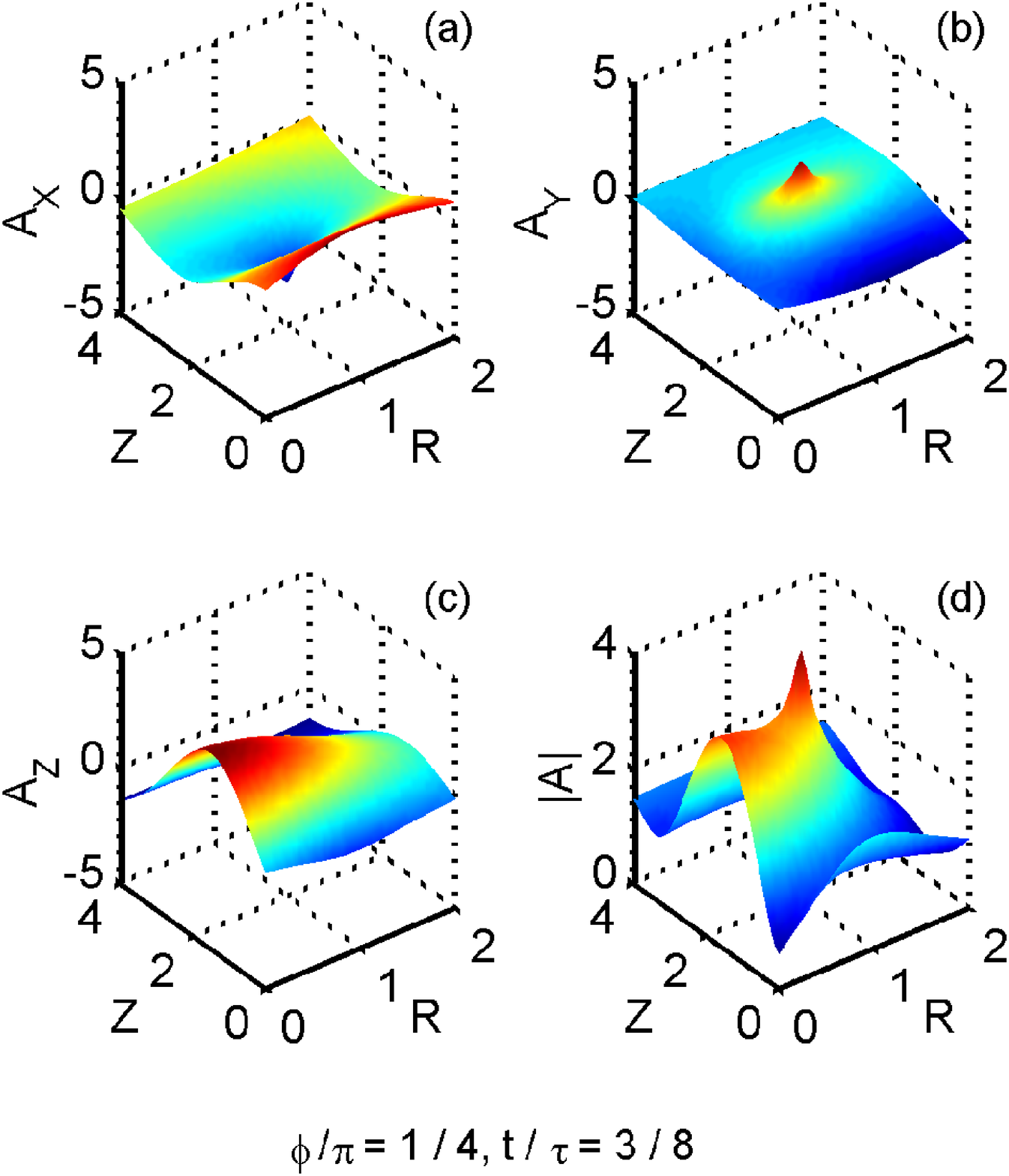}%
\caption{\label{fig:retO} Vector potential for $\nu = 541.6$ MHz at $\phi = \pi / 4$ and $t / \tau = 3/8$.}
\end{figure}

\begin{figure}[]
\includegraphics[width=8.5cm]{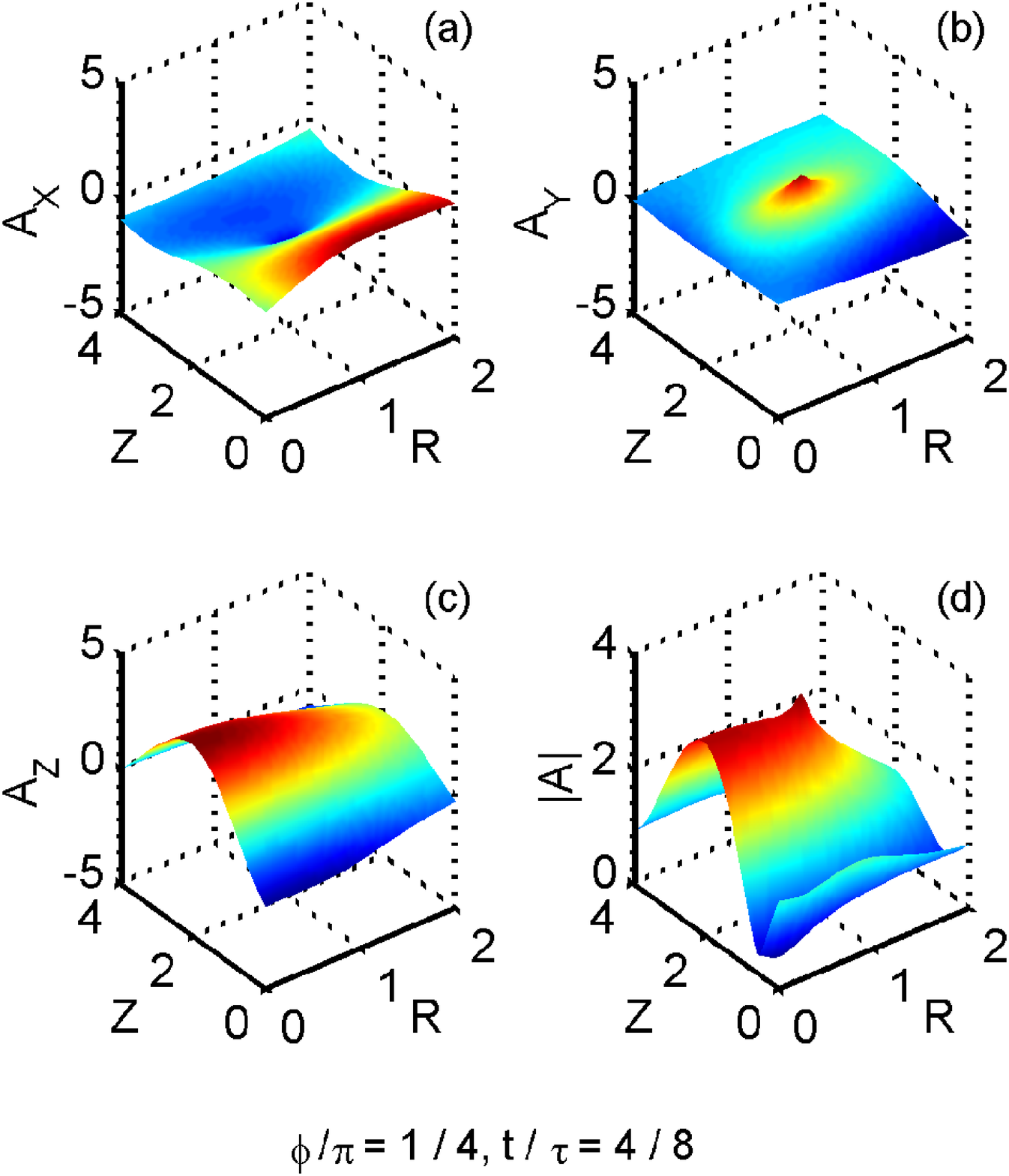}%
\caption{\label{fig:retP} Vector potential for $\nu = 541.6$ MHz at $\phi = \pi / 4$ and $t / \tau = 1/2$.}
\end{figure}

\begin{figure}[]
\includegraphics[width=8.5cm]{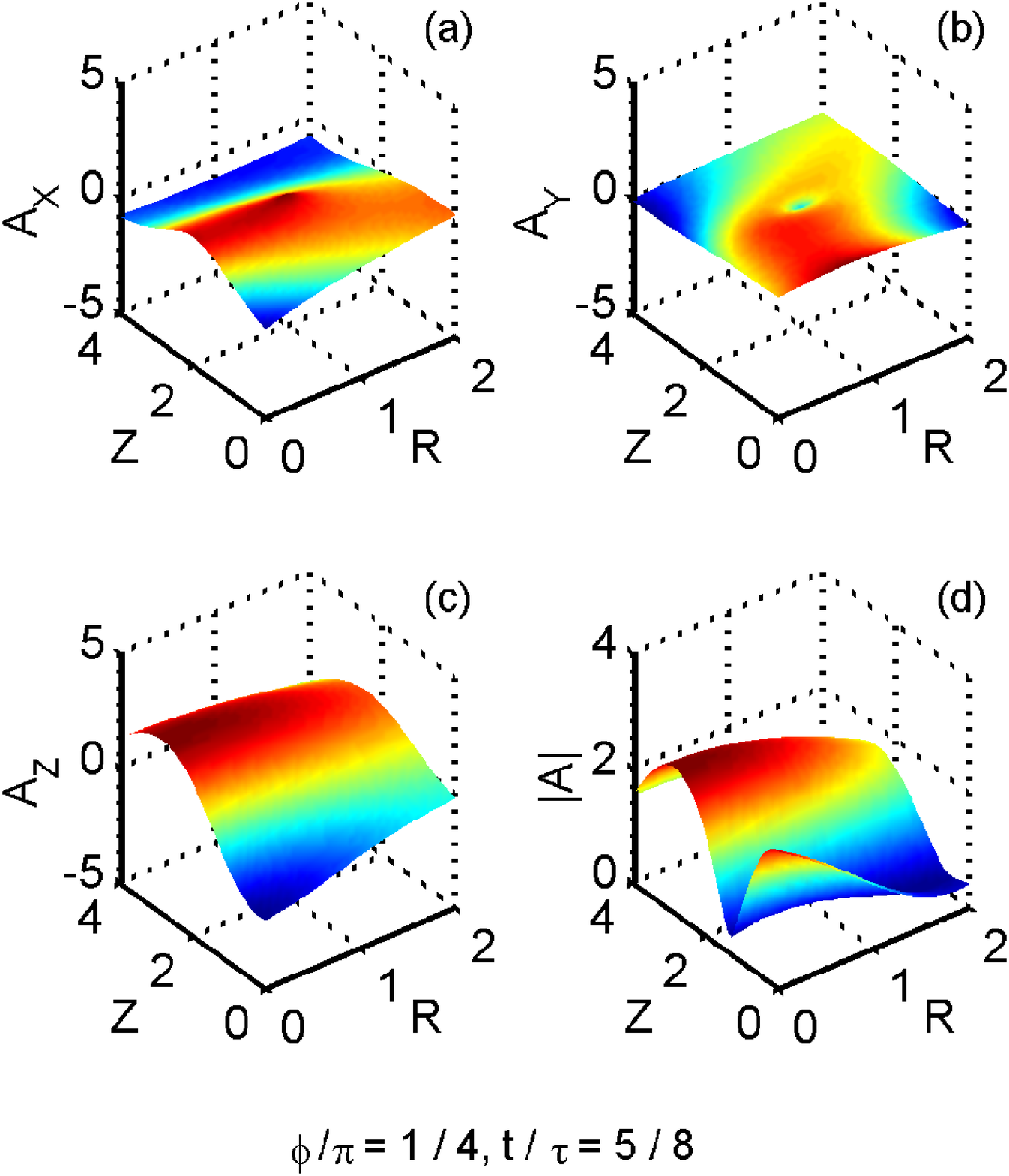}%
\caption{\label{fig:retQ} Vector potential for $\nu = 541.6$ MHz at $\phi = \pi / 4$ and $t / \tau = 5/8$.}
\end{figure}

\begin{figure}[]
\includegraphics[width=8.5cm]{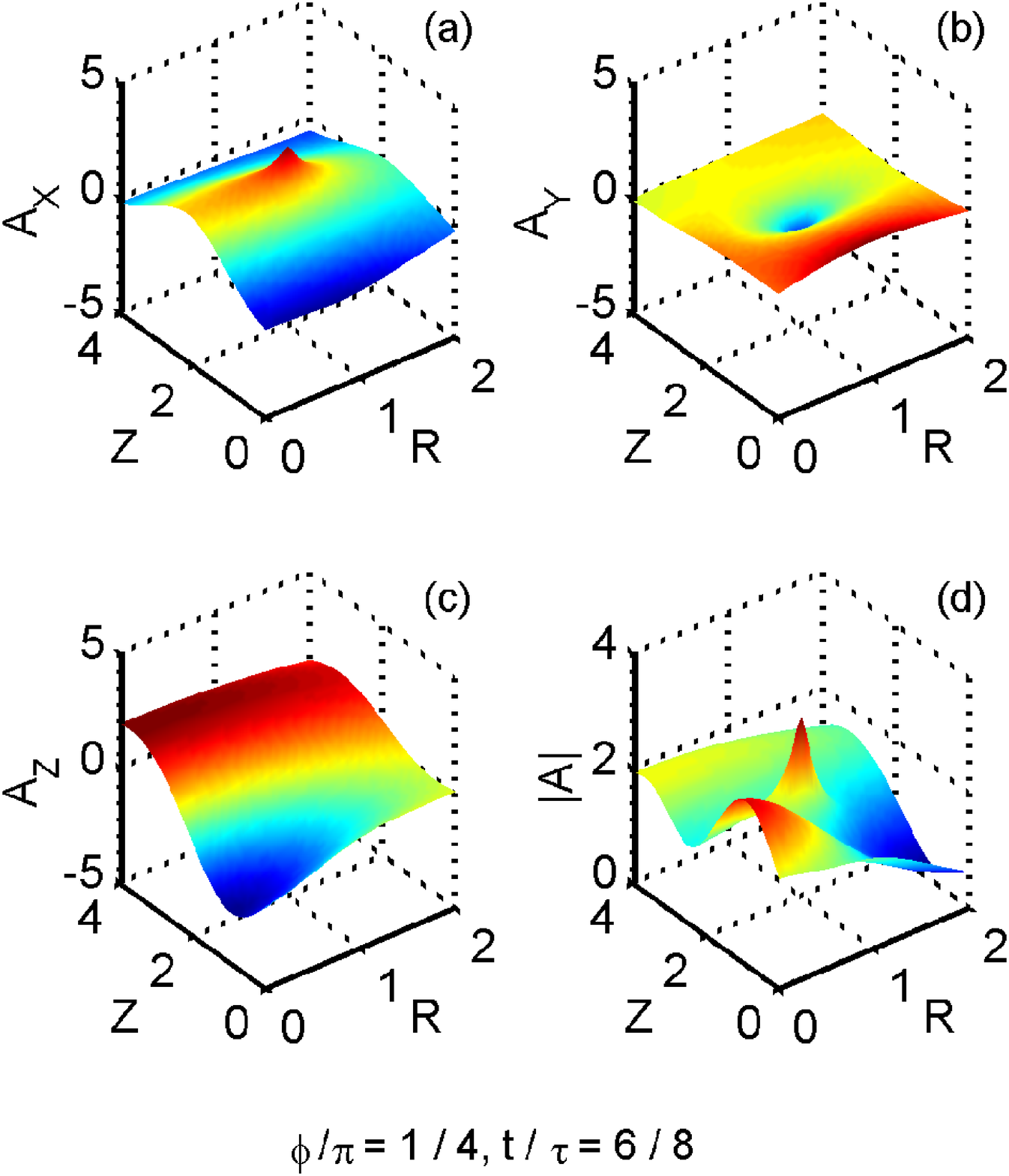}%
\caption{\label{fig:retR} Vector potential for $\nu = 541.6$ MHz at $\phi = \pi / 4$ and $t / \tau = 3/4$.}
\end{figure}

\begin{figure}[]
\includegraphics[width=8.5cm]{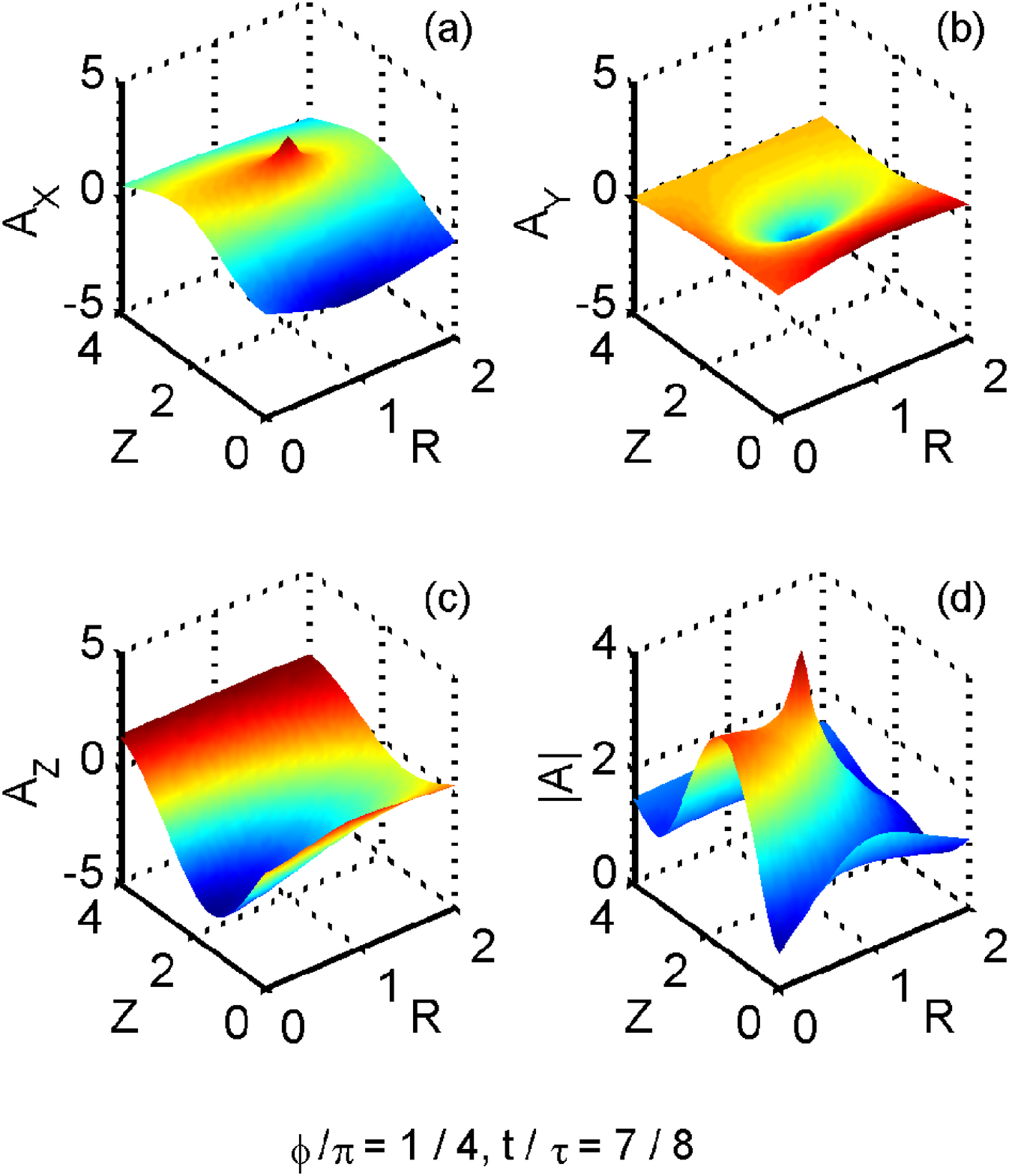}%
\caption{\label{fig:retS} Vector potential for $\nu = 541.6$ MHz at $\phi = \pi / 4$ and $t / \tau = 7/8$.}
\end{figure}

\begin{figure}[]
\includegraphics[width=8.5cm]{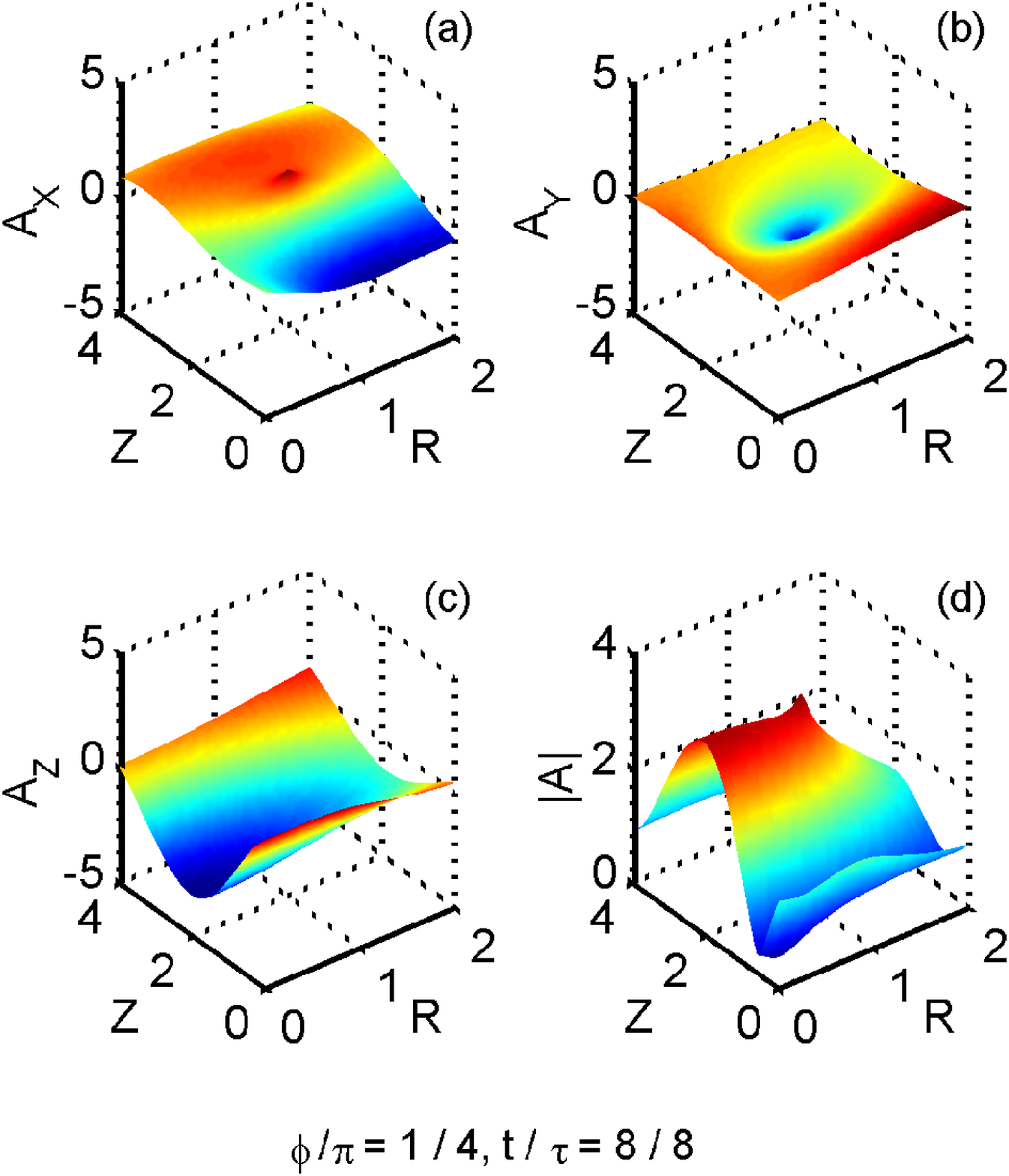}%
\caption{\label{fig:retT} Vector potential for $\nu = 541.6$ MHz at $\phi = \pi / 4$ and $t / \tau = 1$.  Online this figure is linked to a short video showing the temporal evolution of the vector potential.}
\end{figure}

The complicated phase structure along the antenna indicates that we are in a regime in which interference effects can lead to the propagation of an electromagnetic pulse.  Restricting attention to the $ZR$ plane at azimuth $\phi = \pi / 4$, let us now investigate how the pulse evolves through time.  Considering the set of times given by $t / \tau \in [1, 8] / 8$, we display the results in Figs.~\ref{fig:retM} through \ref{fig:retT}, respectively.  Times separated by one half period have the same magnitude for $\mbf{A}$ and vector components of opposite sign.  At the end of one cycle, the potential is back to the configuration found for $\theta_t = 0$, in that Fig.~\ref{fig:retT} is the same as Fig.~\ref{fig:retC}.  What we find most interesting is how a pulse develops in the central region of the antenna, best seen in $\abs{\mbf{A}}$, which then propagates along $\uvec{Z}$.  The two peaks along the $Z$ axis at $R=0$ in Fig.~\ref{fig:retM} panel (d) represent the rectification of the peak and trough of the propagating pulse.  The pulse is seen, by following its development through the figures, to detach from the antenna and continue into space beyond the head ring.  An animation with twice the temporal resolution is available as an online supplement.

\section{Discussion}
The original motivation for this study was simply to develop an interesting exercise in classical electrodynamic field theory, with possible utility to investigators in electric propulsion.  However, by playing around with the driving frequency, we have found evidence that a significant pulse may be generated which propagates away from the antenna.  We expect interesting results when the ratio $\lambda_\mrm{max} / \lambda_0$ is at half-integer multiples, so that constructive interference from the head and tail rings leads to a propagating pulse.  Noticing the similarity between the higher frequency here and the range spanned by existing RF injection plasma heating technologies based on resonance with the ion cyclotron and the lower hybrid frequencies, together spanning a range of about 0.05 to 10 GHz~\cite{dendybook-93}, we wonder if the helicon antenna might be adapted for use in fusion engineering.  For that purpose, the inclusion of a ground plane to achieve unidirectional propagation might be necessary to allow for mounting on the outside of the vacuum vessel, or perhaps the antenna unaltered can be placed towards the interior of the central solenoid in a tokamak so that both propagation pulses can be utilized.

The model here has made several simplifying approximations which need to be more fully developed.  As the material of the conductor is unspecified, we have used the vacuum values of $\epsi_0$ and $\mu_0$ in the determination of the phase along the antenna $\theta_\lambda$.  For greater veracity, the values of $\epsi$ and $\mu$ appropriate for the chosen material should be used instead.  A volume or surface current density could be used for the antenna rather than the line current used here, with straightforward modifications to the evaluation of the integrals.  More interesting but more difficult is the incorporation of self interactions between different parts of the antenna---the current $\mbf{I}_k$ cannot be immune to the effects of the current at other locations $\mbf{I}_{k'}$, which are beyond the scope of this article.  Finally, one is interested in how the propagating electromagnetic pulse would interact with a medium of ionized gas located in the region beyond the antenna rings.  By properly tuning the antenna dimensions and driving frequency, one hopes to drive resonances inside the plasma medium.

\section{Conclusion}
In this article we have presented a numerical evaluation of the vector potential produced by the helicon antenna in vacuum for both the static case and the dynamic case of nonzero driving frequency.  We have considered both a typical driving frequency and one somewhat greater, where the higher frequency produces a significant electromagnetic pulse which propagates along the cylindrical axis.  The frequency of the propagating pulse for typical size parameters lies within the range spanned by ion cyclotron and lower hybrid resonances, suggesting that an adaptation of the helicon antenna design might be useful for RF injection plasma heating.  Extensions of the model for a more realistic description of an experimental apparatus are outlined.



%

\end{document}